\documentclass[prl,aps,showpacs,superscriptaddress]{revtex4}%
\usepackage{amsfonts}
\usepackage{amsmath}
\usepackage{amssymb}
\usepackage{graphics}
\usepackage{mathptmx}

\begin{document}
\title{Energy-Resolved Electron-Spin Dynamics at Surfaces of p-Doped
GaAs} 
\author{H. C. Schneider}
\author{J.-P. W\"{u}stenberg}
\author{O. Andreyev}
\author{K. Hiebbner}
\author{L. Guo}
\altaffiliation{Now at Dept. of Physical Chemistry, Chalmers University, Gothenburg, Sweden}
\author{J. Lange}
\affiliation{Physics Department, Kaiserslautern University, 67653 Kaiserslautern, Germany} 
\author{L. Schreiber} 
\author{B. Beschoten}
\affiliation{II. Physikalisches Institut, RWTH Aachen, %
and Virtual Institute for Spin Electronics (VISel), Templergraben 55, 52056 Aachen, Germany}
\author{M. Bauer}
\author{M. Aeschlimann} 
\affiliation{Physics Department, Kaiserslautern University, 67653 Kaiserslautern, Germany}

\date{\today}


\begin{abstract}
Electron-spin relaxation at different surfaces of p-doped GaAs is
investigated by means of spin, time and energy resolved 2-photon
photoemission. These results are contrasted with bulk results obtained
by time-resolved Faraday rotation measurements as well as calculations
of the Bir-Aronov-Pikus spin-flip mechanism. Due to the reduced hole
density in the band bending region at the (100) surface the
spin-relaxation time increases over two orders of magnitude towards
lower energies. At the flat-band (011) surface a constant spin
relaxation time in agreement with our measurements and
calculations for bulk GaAs is obtained.
\end{abstract}

\maketitle

In recent years, much experimental and theoretical work has been
focused on the control and manipulation of the electronic spin degree
of freedom independently of its charge, with the ultimate goal of
spintronics devices, in which the electron spins are the carriers of
the information~\cite{awschalom:phystoday97}. The limiting factor for
the usefulness of the information encoded in a spin-polarized current
in a non-ferromagnetic semiconductor is the relaxation of the spin
polarization, which is caused by a variety of interaction
mechanisms~\cite{zutic:review}. In bulk GaAs, the relaxation of
optically induced spin polarizations has been studied intensely for
more than 30 years. Early work has led to the identification of
several mechanisms that destroy the spin polarization, and good
agreement between experiment and theory was found on the level of
numerical and experimental accuracy available at that
time~\cite{opt-orient}. In recent years, there has been renewed
experimental and theoretical interest in spin relaxation, with many
experimental studies focusing on undoped and n-doped
semiconductors~\cite{kikkawa:prl98,heberle:prl01}. Early results for
p-doped GaAs were obtained by means of hot photon luminescence and the
Hanle effect~\cite{zerrouati:prb88}. Surfaces and interfaces, such as
Schottky barriers, originally received comparatively little
attention~\cite{pierce:photoemission,riechert:prl84}. More recently,
however, interfaces have been studied because of their importance for
spintronics device applications where efficient electrical spin
injection from a ferromagnetic metal or half-metal through a Schottky
barrier into the semiconductor is of utmost
importance~\cite{zhu:prl01:rt-spin-injection}. The difficulty of
efficient spin injection has stimulated interest in a fundamental
understanding of spin-flip scattering at semiconductor surfaces and
interfaces, e.~g., at step edges~\cite{labella:science01}.

The purpose of this paper is the investigation of electron
spin-relaxation in p-doped GaAs and the unambiguous identification of
surface effects on the electron spin-relaxation. Using a spin, energy
and time-resolved photoemission technique~\cite{ma:prl97} we study the
room-temperature spin-dependent electron dynamics at two surfaces with
different characteristics~\cite{moench:book}: the (100) surface with
pronounced band-bending and the cleaved (011) surface, for which we
expect flat-band conditions. To obtain a complete picture of the spin
relaxation at surfaces as compared to the bulk we have also measured
bulk spin relaxation-times by time-resolved Faraday rotation (TRFR) on
identical samples. Furthermore, we numerically evaluate for the first
time the full momentum-dependent room-temperature bulk spin-relaxation
rate for the Bir-Aronov-Pikus (BAP)~\cite{bap:76} electron spin-flip
mechanism due to the electron-hole exchange interaction.

One should stress that we experimentally and theoretically study the
dynamics of the \emph{incoherent}, energy resolved, microscopic spin
polarization $P= (n_{\uparrow} - n_{\downarrow})/(n_{\uparrow} +
n_{\downarrow})$ defined in terms of the transient, energy or momentum
dependent electron distributions for spin-up ($n_\uparrow$) and
spin-down ($n_\downarrow$) electrons. The dynamics of this microscopic
spin polarization determines the relaxation of the macroscopic spin
polarization, which is often described by a phenomenological
(longitudinal) $T_1$ time. We therefore refer to the decay of the
electron spin-polarization simply as spin relaxation. The experimental
and theoretical results on spin relaxation presented here are obtained
without external magnetic fields, and should therefore not be confused
with the dephasing of \emph{coherent} spin dynamics under the
influence of magnetic fields, whose macroscopic counterpart is a
(transverse) $T_2$ time.

At GaAs surfaces and GaAs/metal interfaces a Schottky barrier exists,
in which the Fermi level pinning causes a band bending downward of
magnitude up to 0.6\,eV in p-doped GaAs~\cite{moench:book}. For the
experimental determination of the spin relaxation in the band bending
region, we use a time-domain pump-probe approach known as
time-resolved 2-photon-photoemission (TR-2PPE) to investigate excited
electron dynamics with femtosecond time resolution. In our setup, a
circularly polarized pump pulse with a photon energy just above the
bulk band-gap [process (1) in Fig.~1] can excite spin polarized
electrons well in the bulk because of its penetration depth of several
100\,nm. A probe pulse with a higher photon energy and smaller
penetration depth (several 10\,nm) removes electrons close to the
surface [process (3)] after a time delay $\tau$, during which the
electrons have traveled towards the surface and undergone both
momentum and spin-flip scattering processes [process (2)]. The kinetic
energy of the photoemitted electrons together with their spin
polarization is directly measured in a spin-sensitive low-energy
electron diffraction (SPLEED) detector~\cite{kirschner-contrib}. One
of the advantages of this technique is that it yields the
energy-resolved spin polarization $P(E,t)$ \emph{independently of the
electron density}. Electronic transitions due to, e.g.,
carrier-carrier scattering, or electron-hole recombination that reduce
the carrier concentration in the investigated energy interval only
result in an increased statistical error in the measured spin
polarization.

\begin{figure}[bt]
\centering \resizebox{0.4\textwidth}{!}{\rotatebox{270}%
{\includegraphics{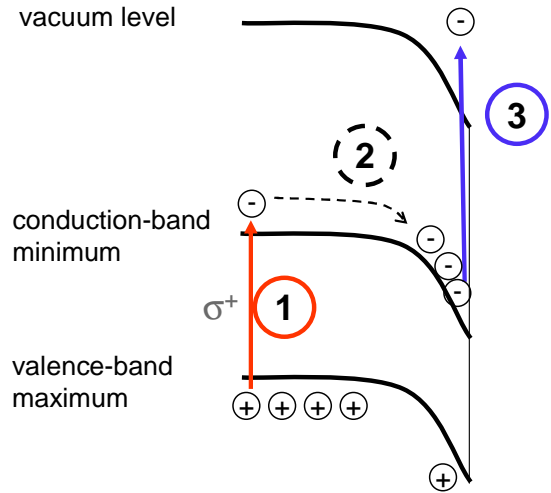}}}
\caption{Two-photon photoemission process at a surface region with
  band bending involving: (1) the excitation of spin-polarized
  electrons by the polarized pump pulse away from the surface, (2) the
  scattering of electrons into low energy states in the band bending
  region, and (3) the photoemission process from the surface due to the
  probe pulse with time delay $\tau$. The hole density at the surface
  is reduced due to the hole-band bending.}
\end{figure}

Our TR-2PPE setup employs a Ti:sapphire laser at 82 MHz that yields
linearly polarized light pulses of 50\,fs FWHM [process (1) in Fig.~1]
with a pump photon energy of 1.55\,eV. A fraction of the light is
frequency-doubled, leading to a probe photon energy of 3.1\,eV
[process (3) in Fig.~1]. The fundamental (pump) pulse is circularly
polarized by a quarter-wave plate whereas the frequency-doubled
(probe) pulse remains linearly p-polarized. The two pulses are
mechanically delayed with respect to each other using a Mach-Zehnder
interferometer. The collinear pulse pair is focused on the sample
surface within a UHV chamber. As sample we use a p-doped GaAs crystal
with doping concentration (acceptor: zinc) $N_{\text{A}}= 5\times
10^{18}\,\text{cm}^{-3}$ both in (100) and (011) orientation. Prior to
the measurements the sample was treated with a small amount of cesium,
thus leading to a well-defined Fermi-level-pinning and a lowered
work-function of about 3.2\,eV. The excited electron densities were
about $10^{16}\,\text{cm}^{-3}$.


%
\begin{figure}[tb]
\centering \resizebox{0.6\textwidth}{!}{\rotatebox{270}%
{\includegraphics{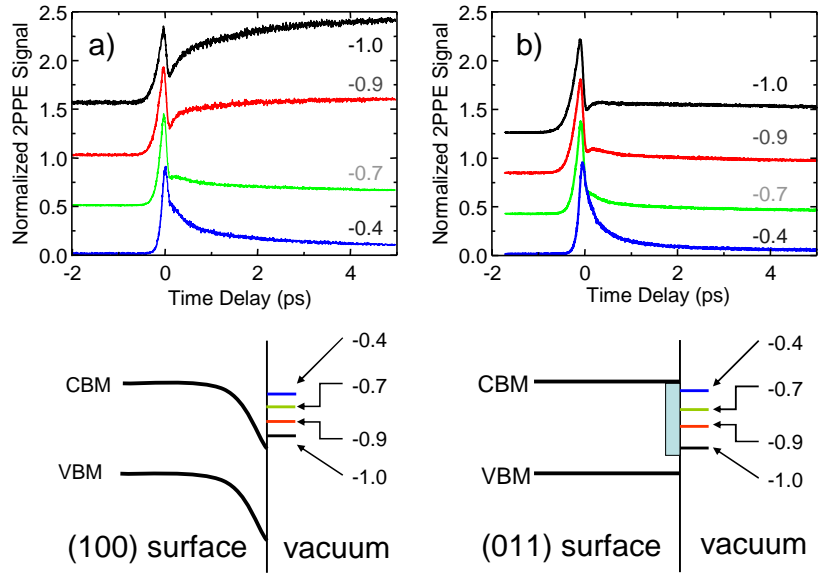}}}
\caption{Normalized 2PPE signal vs.\ time for different electronic
  energies at the GaAs (100) surface (a) and (011) surface (b). The
  corresponding band profiles displaying the conduction-band minimum
  (CBM) and the valence-band maximum (VBM) are sketched in the lower
  panel. Energies indicated in this figure are in units of eV and are
  measured with respect to the bulk CBM.}
\end{figure}

Figure~2 shows normalized spin-integrated photoemission signals, i.e.,
the number of detected electrons at a fixed kinetic energy, for the
GaAs (100) surface (pronounced band bending) and the (011) surface
(essentially flat-band conditions). The electronic energies are
$E-E^0_{\text{cbm}}= -1.0$, $-0.9$, $-0.7$, and $-0.4$\,eV measured
with respect to the bulk conduction-band minimum (CBM)
$E^0_{\text{cbm}}$. A time-independent background due to electrons
photoemitted by multiphoton processes has been subtracted in the
curves in Fig.~2. The spike around zero time delay is due to the
photoemission of unpolarized electrons during the overlap of pump and
probe pulses. For positive time delay $\tau$ the 2PPE signal monitors
the population dynamics at the surface.  In the band bending region of
the (100) surface at $E-E^0_{\text{cbm}}= -1.0$\,eV, the 2PPE signal in
Fig.~2(a) indicates a \emph{rising} electron population over
5\,ps. For the next higher energy in the conduction band this increase
is less pronounced, and for the conduction-band states at still higher
energies the populations decay with a time constant of
5\,ps. This behavior is due to refilling processes that occur when
``hot'' electrons are scattered into states close to the CBM from
higher energies. The population of states at higher energies, e.g.,
$E-E_{\text{cbm}} = -0.4$\,eV decreases due to outscattering processes
whereas the population of states at lower energies, e.g.,
$E-E_{\text{cbm}} \leq -1.0$\,eV is refilled with these
electrons. There is also the possibility of a weakening of the band
bending by the space charge due to the photoexcited
electrons~\cite{deJong:Schottky}. However, this mechanism will only
make the refilling more pronounced when the conduction band bends
downwards in the process of charge equilibration on a timescale of a
few picoseconds after photoexcitation of the carriers.

Figure~2(b) shows the 2PPE measurements at the flat-band (011) surface
that indicate fast electronic dynamics. However, the kinetic energies
of the photoemitted electrons now correspond to electronic surface
states in the fundamental gap. For delays of more than 1\,ps these
states all show a slowly decreasing electron population, in contrast
to the result obtained for the (100) surface.

\begin{figure}[t]
\centering \resizebox{0.4\textwidth}{!}{\rotatebox{270}%
{\includegraphics{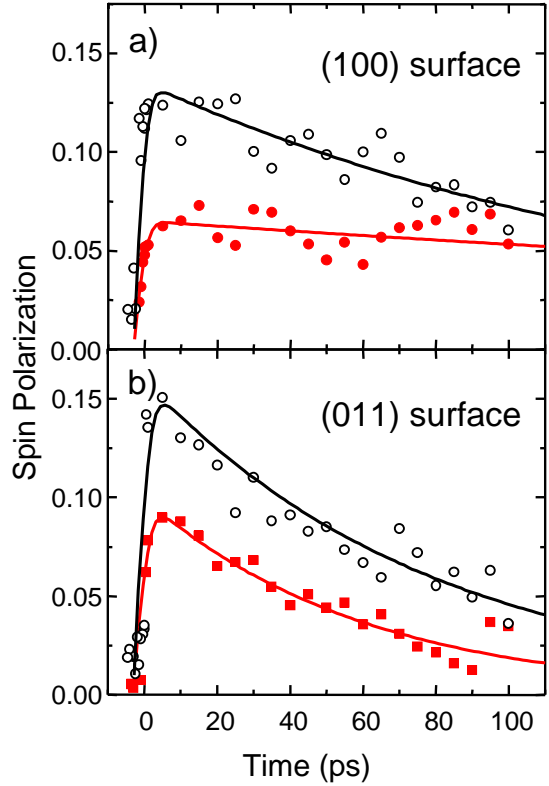}}}
\caption{Dynamical spin polarization (dots) derived from 2PPE signal
  photoemitted from (a) the GaAs (100) surface and (b) the (011)
  surface, as well as corresponding exponential fits.  The electron
  kinetic energies are $E-E^0_\text{cbm}= -1.0$\,eV (upper curves)
  and $E-E^0_\text{cbm}= -1.1$\,eV (lower curves).}
\end{figure}

Figure~3 shows the dynamical spin polarization obtained from the
spin-resolved dynamical 2PPE signals~\cite{kirschner-contrib} for
electrons with definite kinetic energies $E-E^0_\text{cbm}= -1.1$\,eV
and~$-1.0$\,eV from the GaAs (100) and (011) surfaces. Comparing the
delayed rise of the dynamical spin polarization in Fig.~3 to the rise
of the spin-integrated signals in Fig.~2, one notices that carriers
emitted during the overlap of pump and probe pulses are mainly
unpolarized because otherwise the spin-polarization should reach its
maximum at the same time as the spin-integrated signal. By fitting the
polarization dynamics to an exponential decay for delay times $\tau >
10$\,ps, we obtain spin-relaxation times for states at the surface
over the whole energy range of the band bending. For the (100) surface
and energies $E-E^0_\text{cbm}= -1.1$\,eV and $-1.0$\,eV, we obtain
spin-relaxation times of $\tau_{\text{spin}} \approx 500$\,ps and
150\,ps, respectively. In the case of the (011) surface without
band-bending, we obtain from Fig.~3(b) at the same energies
$E-E^0_\mathrm{cbm}= -1.1$\,eV and $-1.0$\,eV almost constant spin
relaxation times of 60\,ps and 70\,ps, respectively. It should be
stressed that the spin-resolved 2PPE method allows us to monitor the
spin polarization up to more than 100 picoseconds when the
photoelectron yield, and thus the carrier density, is very small.

Figure~4 shows the spin-relaxation times extracted from the spin- and
energy-resolved 2PPE measurements over a range of energies for both
the GaAs (100) and (011) surfaces. For the GaAs (100) surface we
observe an increase from about 60\,ps to 1600\,ps over an energy range
of more than 0.3\,eV, whereas the spin-relaxation time for the (011)
surface is about 60\,ps, independent of energy. The bulk spin
relaxation was measured on an identical sample by means of the
time-resolved magneto-optical Faraday
effect~\cite{crooker:prb97,beschoten:gan-spincoherence:prb01}. The
bulk spin-relaxation time was found to be 60\,ps at room temperature
regardless of the crystal orientation and is also shown in Figure~4 as
a guide to the eye. The spin-relaxation times for the (011) surface
and at higher energies for the (100) surface are equal to the bulk
value. In these cases there appear to be no additional contributions
to electron spin-flip scattering that enhance the spin relaxation.

It remains to explain the different energy dependeces for both
surfaces in the light of earlier theoretical results that predict an
increasing spin relaxation time at lower kinetic energies due to
\emph{intrinsic} properties of the BAP
mechanism~\cite{maialle:prb96}. To clarify this point we theoretically
investigate the spin-relaxation time due to the BAP
mechanism~\cite{opt-orient}, which is expected to dominate in the
present temperature and doping density range~\cite{song:prb02}. The
original analysis~\cite{bap:76} introduced an explicitly momentum
dependent spin decay rate $1/(2\tau^s_\mathrm{BAP})$ in Born
approximation that describes the spin-dependent population relaxation
$\partial_{t}n_{sk} = -1/(2\tau^s_\mathrm{BAP}(k)) [n_{s,k}-f_{s,k}]$
with
\begin{equation}
  \frac{1}{2\tau^s_\mathrm{BAP}(k)} = \frac{2\pi}{\hbar}
  \sum_{\vec{q},\vec{p}}\sum_{j,j'}
  |\langle j's'|V_{\text{exc}}(\vec{q})|sj\rangle|n_{j,p}(1-n_{j,\vec{p}-\vec{q}})
   \delta(\epsilon_{j,p}+\epsilon_{s,k}-\epsilon_{-s,\vec{k}+\vec{q}}-
                \epsilon_{j',\vec{p}-\vec{q}})
\label{BAP-rate}
\end{equation}
Here, $n_{s(j),k}$ are the momentum dependent carrier distributions,
$s$ is the spin projection quantum number and $j=\pm3/2$, $\pm1/2$ the
hole angular momentum projection quantum number. The electron and hole
energies are denoted by $\epsilon_s$ and $\epsilon_j$,
respectively. The interaction matrix element due to long-range and
short-range exchange interaction $\langle
j's'|V_{\text{exc}}(\vec{q})|sj\rangle$ is defined in
Ref.~\cite{maialle:prb96}. The electron equilibrium distribution,
towards which the electron spin relaxes, is denoted by $f_{s,k}$.
Using this relaxation time equation and taking into account that an
electron that undergoes a spin-flip increases the number of electrons
with opposite spin, one obtains for the case of unpolarized holes that
the spin decay rates $1/(2\tau^s_\text{BAP})$ are independent of the
spin orientation $s$ and consequently $\partial_t P_k(t)=
-[\tau_{\text{BAP}}(k)]^{-1}\,P_k(t)$ as the dynamical equation for
the momentum dependent electron spin polarization $P_k$.  Using this
reasoning the spin decay rate has been computed as a measure of the
spin relaxation~\cite{maialle:prb96} and was found to be strongly
momentum dependent for both bulk and quantum well GaAs at
$T=0$\,K. Our explicit numerical evaluation of Eq.~(\ref{BAP-rate})
without further approximations shows that this momentum dependence,
which translates into an energy dependence, of the spin lifetime
persists at room temperature. However, for low electron densities
around room temperature as in the case of our experiments, the
electronic distributions $n_{sk}$ can be approximated by
Maxwell-Boltzmann distributions so that $P_k(t)$ becomes
\emph{momentum (energy) independent} and the spin relaxation rate is
given by the energy independent average of Eq.~(\ref{BAP-rate}). This
result has been checked against a numerical solution of the full
Boltzmann equation for bulk GaAs including both spin-flip exchange
scattering and spin-conserving direct Coulomb scattering similar to
the approach in Ref.~\cite{wu:prb00:kinetics_qw}. The inset in Fig.~4
shows that the bulk results calculated in this way for a hole density
of $5\times 10^{18}\,\text{cm}^{-3}$ are in good qualitative agreement
with the Faraday effect experiments over a temperature range of more
than 100\,K, in which the BAP mechanism is expected to be the dominant
spin-relaxation mechanism. Since the calculated spin relaxation due to
the BAP mechanism in our experiments is energy independent, the energy
dependence of the spin relaxation-times in the band-bending region of
the GaAs (100) surface can only be explained by a change of the
material properties at the surface, which modifies the BAP scattering
efficiency for electrons localized by the band-bending potential at
the surface. This is most likely the reduced density of holes in the
band bending region since the holes are driven away from the
surface. Using our calculations, we estimate that a reduction of the
hole doping concentration of about one order of magnitude in the band
bending region is responsible for the long lived spin polarization at
the (100) surface. The above argument applies to low-energy electrons
localized at the (100) surface; electrons at higher energies are not
localized in the band bending region and therefore show the same
energy independent spin relaxation time as electrons at the (011)
surface and in the bulk.

\begin{figure}[t]
\centering
\resizebox{0.5\textwidth}{!}{\rotatebox{270}{\includegraphics{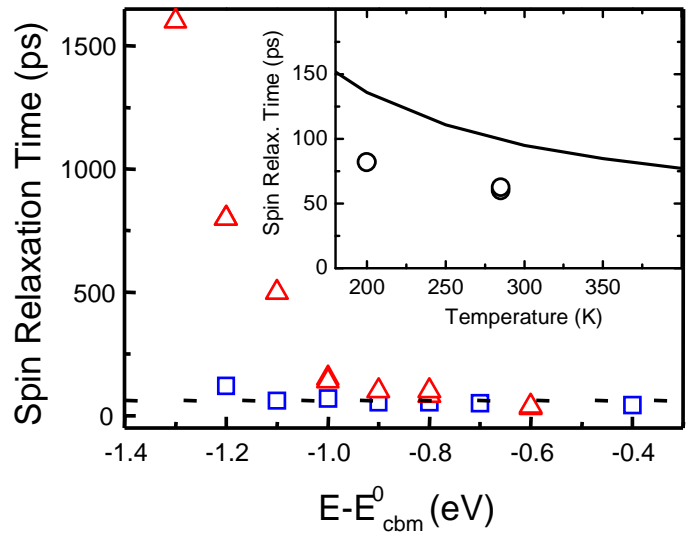}}}
\caption{Spin relaxation times vs.\ energy for the GaAs (100) surface
  (triangles) and (011) surface (squares). The dashed line indicates
  the bulk spin relaxation-time at room temperature obtained from TRFR
  measurements. Inset: Temperature dependence of \emph{bulk} spin
  relaxation time obtained from TRFR measurements (open circles) and
  calculations using the BAP mechanism (solid line).}
\end{figure}

In summary, we have presented a study of spin-flip processes in the
band-bending region of an interface between a p-doped semiconductor
and a metal by means of spin and time-resolved 2PPE.  This method is
complementary to bulk sensitive Faraday rotation measurements, and
yields the energy-resolved spin dynamics of electrons at surfaces. It
is shown that the spin-relaxation time can exceed the carrier lifetime by
an order of magnitude. Comparing these results to spin
relaxation times in the bulk as obtained from our Faraday rotation
measurements on identical samples as well as the numerical evaluation
of the BAP spin-flip scattering rate, we do not find additional
contributions for spin scattering at the (011) surface. For the (100)
surface, the spin relaxation-rate is decreased compared to the bulk
value for electrons in the band-bending region at the surface due to
the lower concentration of holes, which act as scattering partners for
the spin-flip electron-hole exchange scattering.

\begin{acknowledgments}
We thank M.~Fleischhauer, W.~H\"ubner, and S.~W.~Koch for helpful
discussions. 
A grant for CPU time from the Forschungszentrum J\"ulich
is gratefully acknowledged. This work was supported by the DFG, BMBF,
and HGF.
\end{acknowledgments}

\end{document}